\documentstyle [epsfig,emulateapj] {article}

\newcommand{\be}{\begin{eqnarray}}
\newcommand{\ee}{\end{eqnarray}}
\def\lsim{\mathrel{\rlap{\lower3pt\hbox{\hskip1pt$\sim$}}
	\raise1pt\hbox{$<$}}} 
\def\gsim{\mathrel{\rlap{\lower3pt\hbox{\hskip1pt$\sim$}}
	\raise1pt\hbox{$>$}}} 
\newcommand{\msun}{\mbox{~$M_\odot$}}

\lefthead{Brown, Lee, Lee, Wijers, \& Bethe}
\righthead{Black Holes, Gamma Ray Bursters, Gravitational Mergers}

\begin{document}

\title{Brightness from the Blackest Night: \\
       Bursts of Gamma Rays and Gravity Waves from Black Hole Binaries}
\author{G. E. Brown$^1$, R. A. M. J. Wijers$^1$, C.-H. Lee$^1$, H. K. Lee$^2$,
        and H. A. Bethe$^3$}
\affil{1) Department of Physics \& Astronomy,
        State University of New York,
        Stony Brook, New York 11794, USA\\
       2) Department of Physics, Hanyang University, Seoul 133-791, Korea\\
       3) Floyd R. Newman Laboratory of Nuclear Studies,
        Cornell University, Ithaca, New York 14853, USA}


\begin{abstract}
We use recent results in binary stellar evolution to argue that
binaries with at least one black hole dominate the rate of compact-object
mergers. Two phenomena generally attributable to such mergers,
gamma-ray bursts and gravity-wave bursts, are therefore likely to
originate from near the event horizon of a black hole.
In addition to sheer numbers, black holes have an added advantage
over neutron stars in both phenomena. For gamma-ray bursts, the presence
of an event horizon eases the baryon pollution problem, because energy
can be stored into rotation until most baryons have been swallowed, and then
released into a cleaner environment via the Blandford-Znajek process.
For gravity-wave bursts, black holes offer higher luminosities due to
their higher masses, thus enabling detection out to larger distances,
which leads to a 30-fold increase in the predicted LIGO event rate.
\end{abstract}

\section{Introduction}
\label{sec1}

Binaries containing a black hole, or single black holes, have been
suggested for some time as good progenitors for gamma-ray bursts
(Paczy\'nski 1991, 1998, Mochkovitch et~al.\ 1993,
Woosley 1993, Fryer \& Woosley 1998, MacFadyen
\& Woosley 1998). Reasons for this include
the fact that the rest mass of a stellar mass black hole is comparable
to what is required to energize the strongest GRB. Also, the horizon
of a black hole provides a way of quickly removing most of the material
present in the cataclysmic event that formed it. This may be important
because of the baryon pollution problem: we need the ejecta that give
rise to the GRB to be accelerated to a Lorentz factor of 100 or more,
whereas the natural energy scale for any particle near a black hole is
less than its mass. Consequently, we have a distillation problem of taking
all the energy released and putting it into a small fraction of the total
mass. The use of a Poynting flux from a black hole in a magnetic field
(Blandford \& Znajek 1977) does not require the presence of much mass,
and uses the rotation energy of the black hole, so it provides naturally
clean power. 

In this paper, we discuss and combine a number of new developments in this
area. First, the population synthesis calculations of Bethe \& Brown
(1998) provide good estimates of the formation rates of various 
suggested GRB progenitors. They stress the importance of black holes of
relatively low mass ($\sim2.4\msun$). Binaries with one neutron star and
one such black hole are ten times more common than NS-NS binaries, and
thus contribute much more to GRB and gravity wave rates. Second,
three of us have recently reviewed the Blandford-Znajek (1977) mechanism
as a possible central engine for GRBs (Lee, Wijers, \& Brown 1999).
We confirm that the basic mechanism works effectively, addressing the
criticism of many authors. 

In section~\ref{src} we discuss the various possible progenitors of GRBs
and their potential for generating the right energy on the right time scale.
In section~\ref{pop} we discuss the formation rate of each of these.
Then we combine these pieces of information to obtain estimates of the
detection rates of GRBs and LIGO-detectable mergers (section~\ref{rate}),
and summarize our findings (section~\ref{con}).

   \section{Stellar sources of gravity waves and gamma-ray bursts}
   \label{src}

When a black hole forms from a single star, as in the collapsar model of
MacFadyen \& Woosley (1998) and the hypernova scenario by Paczy\'nski
(1998) it is surrounded by a substantial
stellar envelope, giving two potential GRB energy sources.
First, accretion can release neutrinos in
such large amounts that $\nu\bar{\nu}$ annihilation produces up to
10$^{52}$\,erg in a pair fireball.  Second, the very large rotation
energy of the black hole can be extracted via the Blandford-Znajek
mechanism if the surrounding matter carries a magnetic field.

Mergers of compact-object binaries are strong sources of gravity waves.
The merger leaves a central compact object that is most likely a black
hole, because it contains more than the maximum mass of a neutron
star. Now little mass is left as surrounding debris, perhaps at most
0.1\msun.  Both accretion and rotation energy are available, but due to
the small ambient mass the accretion energy is less likely  to suffice for
a strong GRB in this case.

   \subsection{The Blandford-Znajek mechanism}
   \label{src.bz}

When a rapidly rotating black hole is immersed in a magnetic field, frame
dragging twists the field lines near the hole, which 
causes a Poynting flux to be emitted from near the black hole. This is
the Blandford-Znajek (1977) mechanism. The source of energy for the flux
is the rotation of the black hole. The source of the field is the surrounding
accretion disk or debris torus. We showed (Lee, Wijers, \& Brown 1999)
that at most 9\% of the rest mass of a rotating black hole can be 
converted to a Poynting flux, making the available energy for powering
a GRB
\be
   E_{{\rm BZ}}=1.6\times 10^{53}\; (M/M_\odot)\; {\rm erg}.
\label{eq:ebz}
\ee
The power depends on the applied magnetic field:
\be
   P_{{\rm BZ}}\sim 6.7\times 10^{50} B_{15}^2
      (M/M_\odot)^2 {\rm erg\; s}^{-1}
\label{eq:pbz}
\ee
(where $B_{15}=B/10^{15}$\,G). This shows that modest variations in the
applied magnetic field may explain a wide range of GRB powers, and therefore
of GRB durations. There has been some recent dispute in the literature
whether this mechanism can indeed be efficient (Li 1999) and whether the
power of the BH is ever significant relative to that from the disk
(Livio, Ogilvie, \& Pringle 1999). The answer in both cases is yes,
as discussed by Lee, Wijers, \& Brown (1999). 

The issue, therefore, in finding efficient GRB sources among black holes
is to find those that spin rapidly. There are a variety of reasons why
a black hole might have high angular momentum. It may have formed from
a rapidly rotating star, so the angular momentum was there all along
(`original spin', according to Blandford 1999); it may also have accreted
angular momentum by interaction with a disk (`venial spin') or have
formed by coalescence of a compact binary (`mortal spin').  We shall
review some of the specific situations that have been proposed in turn.

   \subsection{NS-NS and NS-BH binaries}
   \label{src.nsnsbh}

Neutron star mergers are among the oldest proposed cosmological GRB sources 
(Eichler et~al.\ 1989, Goodman, Dar, \& Nussinov 1987, Paczy\'nski 1986), 
and especially the neutrino flux is still actively studied as a GRB power
source (see, e.g., Ruffert \& Janka 1998). 
However, once the central mass has collapsed to
a black hole it becomes a good source for BZ power, since it naturally
spins rapidly due to inheritance of angular momentum from the binary
(Rees \& M\'esz\'aros 1992). Likewise BH-NS binaries (Lattimer \& Schramm
1974) will rapidly transfer a
large amount of mass once the NS fills its Roche lobe, giving a rapidly
rotating BH (Kluzniak \& Lee 1998). The NS remnant may then
be tidally destroyed, leading to a compact torus around the BH. It
is unlikely that this would be long-lived enough to produce the longer
GRB, but perhaps the short ($t\lsim1$\,s) ones could be produced (e.g.,
Fryer, Woosley \& Hartmann 1999). However, mass transfer could stabilize
and lead to a widening binary in which the NS lives until its mass drops
to the minimum mass of about 0.1\msun, and then becomes a debris torus
(Portegies Zwart 1998). By then, it is far enough away that the resulting
disk life time exceeds 1000\,s, allowing even the longer GRB to be made.
Thus BH-NS and NS-NS binaries are quite promising. They have the added
advantage that their environment is naturally reasonably clean, since
there is no stellar envelope, and much of the initially present baryonic
material vanishes into the horizon.

   \subsection{Wolf-Rayet stars}
   \label{src.WR}

The formation of a black hole directly out of a massive star has
been considered for the production of GRB, either as hypernovae
(Paczy\'nski 1998), failed supernovae (Woosley 1993) or exploding WR
stars (MacFadyen \& Woosley 1999).

Another significant source of such events is the formation of a BH of
about 7\msun\ in black hole transients, which is discussed by Brown, Lee,
\& Bethe (1999).  These BHs form from a helium star, because spiral-in
of the companion has stripped the primary of its envelope.

Both the above scenarios suffer from a problem found by Spruit \&
Phinney (1998) with rotation of neutron stars: magnetic fields grown by
differential rotation in the star may efficiently couple the core and
envelope, prohibiting the core to ever rotate rapidly. Then the black
holes formed in the above two ways would not contain enough spin energy
to power a GRB, leaving only the more limited $\nu\bar{\nu}$ energy.

A third variety of black hole in a WR star would come from BH-WR mergers
(Fryer \& Woosley 1998). These happen in the same kinds of systems that
form BH-NS binaries as discussed above, in cases where the initial separation
is smaller, so that spiral-in leads to complete merger rather than
formation of a binary. In this case, the BH and WR star are both spun up
during the spiral-in process (i.e., part of the orbital angular momentum
of the binary becomes spin angular momentum). Then there is enough
spin in the system to power a GRB via the Blandford-Znajek process.

   \section{Progenitor formation rates}
   \label{pop}

In order to evaluate the birth rates of the various progenitors
discussed above, we need to establish the evolutionary paths from
initial binaries taken by each, and then compute the fraction of all
ZAMS binaries that evolve into the desired system. Such a population
synthesis calculation is often done with large Monte Carlo codes (e.g.\
Portegies Zwart \& Yungelson 1998). Here we follow the
treatment by Bethe and Brown (1998), because it is analytic and thus
it is relatively transparent how the results depend on the initial
assumptions. It is limited to systems in which at least one star is
massive enough to produce a supernova.
Their final numbers agree remarkably well with the Monte
Carlo simulations by Portegies Zwart \& Yungelson (1998), if the same 
assumptions about stellar evolution are used in both.

To normalize their rates, Bethe \& Brown (1998) used a supernova rate
of $\alpha=$0.02/yr per galaxy, and assumed that this equaled the birth
rate of stars with mass greater than $10\msun$.  The birth rate of stars
more massive than $M$ scales as $M^{-n}$. Therefore, the supernova rate
in mass interval d$M$ is

\be
   {\rm d}\alpha =\alpha\; n \left(\frac{M}{10\msun}\right)^{-n}
             \frac{{\rm d}M}{M}.
   \label{eq.7}
\ee
In their analysis, Bethe \& Brown use $n=1.5$. Half of all
stars are taken to be close binary systems with separations, $a$, in the
range $0.04-4\times10^{13}$\,cm. The distribution of binary separations
within this range is taken to be flat in $\ln a$. The distribution of
mass ratios, $q$, in binaries with massive primaries is uncertain,
especially at small mass ratios, and we here follow Bethe \& Brown
by taking it to be flat in $q$. All these assumptions, as well as the
details of the evolution scenarios, introduce some amount of uncertainty,
but the good agreement between recent analytic and numerical work suggests
that the formation rates we quote below can be trusted to a factor few.
The results of the discussion on birth rates are summarized in table 1.

%

   \subsection{NS-NS and NS-BH binaries}
   \label{pop.nsnsbh}

In the population synthesis of
Bethe \& Brown (1998), the formation rate of NS-NS binaries comes
out to be $10^{-5}$ per year in the Galaxy, or 10 GEM (Galactic Events
per Megayear). This rate is considerably lower than estimates from
population synthesis calculations prior to Bethe \& Brown (1998) and
Portegies Zwart \& Yungelson (1998), but in good agreement with the
estimated merger rate from the observed neutron star binaries
(Phinney 1991, Van den Heuvel \& Lorimer 1996). The discrepancy between
the older theoretical estimates and newer ones is due to a few factors:
some earlier studies did not include kick velocities, and none included
the destruction of neutron stars by hypercritical accretion. This last
process is an important difference between the Bethe \& Brown analysis
and previous work: they argued that when a neutron star spirals into
a red giant, it accretes matter at a very high rate of up to
1\msun/yr. Then photons are trapped in the flow and the flow cools by
neutrino emission, hence the Eddington limit does not apply. As a result,
the neutron star accretes such a large amount of mass that it exceeds the
maximum mass and turns into a low-mass (2--2.5\msun) black hole. Since the
spiral-in is an essential part of the usual scenario for forming binary
neutron stars, the formation rate is cut down greatly. Only those 
binaries in which the stars initially differ by less than 5\% 
in mass does a binary neutron star form. This
is because in those cases the evolutionary time scales of the
two stars are so close that the initial secondary becomes a giant and 
engulfs the primary when the primary has not yet exploded as a supernova.
Briefly, a close binary of two helium stars exists, and then both
explode as supernovae, disrupting about half the systems.

An immediate consequence of this scenario is that the formation rate
for binaries consisting of a neutron star and a low-mass black hole is
an order of magnitude more, 100 GEM, because this is the fate of all the
systems which in the absence of hypercritical accretion would have become
binary neutron stars. The sum of the formation rates of NS-NS and NS-BH
binaries in the Bethe-Brown scenario is therefore about equal to the NS-NS
formation rate in older studies, providing all other assumptions are the
same. The chief reason why such BH-NS binaries are not seen is the same as
why we generally see only one neutron star of the pair in a NS-NS binary:
the first-born neutron star gets recycled due to the accretion flow from
its companion. If its magnetic field is reduced by a factor 100, as we
observe, its visible lifetime is lengthened by that same factor 100,
since it scales as the inverse of the field strength.  
The second-born pulsar is not recycled, hence only
visible for a few million years and 100 times less likely to be seen. In
BH-NS binaries, the neutron star is the second-born compact object, hence
unrecycled and short-lived. With a ten times higher birth rate but 100
times shorter visible life, one expects to see ten times fewer of them,
and thus the fact that none have yet been seen is understandable.

   \subsection{Wolf-Rayet stars}
   \label{pop.WR}

The rate at which the various progenitors involving WR stars discussed
above (Sect.~\ref{src.WR}) are formed can be calculated easily from the
Bethe \& Brown (1998, 1999) model in the same way they calculated the 
merger rates. 

Helium stars (WR stars) with a low-mass black hole (LBH) in them are
formed from almost the same binaries that make LBH plus NS systems;
the only difference is that they come from smaller initial orbits,
in which the spiral-in does not succeed in ejecting the companion
envelope and thus goes on to the center. From the total available
range in orbital separations, $0.04<a_{13}<4$, LBH-NS binaries are
only made when $0.5<a_{13}<1.9$ (where $a_{13}$ is the separation in
units of $10^{13}$\,cm).  Inside that range, for $0.04<a_{13}<0.5$, the
LBH coalesces with the He core. Hence, using a separation distribution
flat in $\ln a$, coalescences are more common than LBH-NS binaries by a
factor $\ln(0.5/0.04)/\ln(1.9/0.5)=1.9$.  In Bethe \& Brown (1998) the
He star compact object binary was disrupted $\sim 50\%$ of the time in
the last explosion, which we do not have here. Thus, the rate of LBH,
He-star mergers is $3.8$ times the formation rate of LBH-NS
binaries which merge, or $R=3.8\times 10^{-4}{\rm
yr}^{-1}$ in the Galaxy, i.e.\ 380\,GEM.

Bethe \& Brown (1999) found that single stars need to have a ZAMS
mass of at least 80\msun\ to directly form a massive BH,
based on evolution calculations by Woosley, Langer and Weaver (1993).
It is now understood that their He-star mass loss rates were a factor of
at least 2 too high. Calculations with lower mass loss rates carried out
by Wellstein \& Langer (1999) give somewhat higher He-star \& CO core
masses. The further evolution of the CO core has not been calculated
yet, but may lower somewhat the Bethe \& Brown mass limit for high-mass
black-hole formation. Staying with the 80\msun--100\msun\ range for
this route, the rate is $2.5\times10^{-4}{\rm yr}^{-1}$ in the Galaxy.

In addition, we consider the formation of massive black holes of
about 7\msun\ that are seen in soft X-ray transients like A\,0620$-$00.
Their evolution was discussed by Brown, Lee, \& Bethe (1999), who
found a formation rate of $9\times10^{-5}{\rm yr}^{-1}$ in the Galaxy.

   \section{Observable rates}
   \label{rate}

   \subsection{Binary Mergers for LIGO}
   \label{rate.gw}

The combination of masses that will be well determined by LIGO is
the chirp mass
   \be
   M_{{\rm chirp}}=\mu^{3/5} M^{2/5}=(M_1 M_2)^{3/5}(M_1+M_2)^{-1/5}.
   \label{eq.16}
   \ee
The chirp mass of a NS-NS binary, with both neutron stars of mass
$1.4\msun$, is 1.2\,\msun.  A birth rate of 10\,GEM implies a rate of
3 yr$^{-1}$ out to 200\,Mpc (Phinney 1991).  Kip Thorne informs us that
LIGO's first long gravitational-wave search in 2002$-$2003 as discussed
for binary neutron stars is expected to see binaries with $M_{{\rm
chirp}}=1.2\msun$ out to 21\,Mpc.

The chirp mass corresponding to the Bethe \& Brown (1998) LBH-NS
binary with masses $2.4\msun$ and $1.4\msun$, respectively, is
1.6\msun. Including a $\sim 30\%$ increase in the rate to allow for
high-mass black-hole (HBH)-NS mergers (Bethe \& Brown 1999) gives
a 26 times higher rate than Phinney's estimate for NS-NS mergers
($10^{-5}$ yr$^{-1}$ in the Galaxy).  These factors are calculated
from the signal to noise ratio, which goes as $M_{{\rm chirp}}^{5/6}$,
and then cubing it to obtain the volume of detectability, which is
therefore proportional to $M_{{\rm chirp}}^{5/2}$. We then predict
a rate of $3\times(21/200)^3\times26=0.09$ yr$^{-1}$.  This rate is
slim for 2003. The enhanced LIGO interferometer planned to begin in
2004 should reach out beyond 150\,Mpc for $M_{{\rm chirp}}=1.2\msun$,
increasing the detection rate to
$3\times(150/200)^3\times 26 =33$ yr$^{-1}$, and HBH-NS mergers used in
these estimates should be considered a lower limit (Sect.~\ref{pop.WR}).
We therefore find that inclusion of black holes in the estimates for
LIGO predict that we will see more mergers per month than NS-NS mergers
per year.

      \subsection{Gamma-ray bursts}
      \label{rate.grb}

Because gamma-ray bursts have a median redshift of 1.5--2 (e.g.\ Wijers
et~al.\ 1998), and the supernova rate at that redshift was 10--20 times
higher than now, the gamma-ray burst rate as observed is higher than
one expects using the above rates. However, for ease of comparison
with evolutionary scenarios we shall use the GRB rate at the present
time (redshift 0) of about 0.1\,GEM. (Wijers et~al.\ (1998) found a
factor 3 lower rate, but had slightly underestimated it because they
overestimated the mean GRB redshift; see Fryer, Woosley, \& Hartmann (1999)
for more extensive discussions of the redshift dependence).
An important uncertainty is the
beaming of gamma-ray bursts: the gamma rays may only be emitted in narrow
cones around the spin axis of the black hole, and therefore most GRBs may
not be seen by us. An upper limit to the ratio of undetected to detected
GRB is 600 (M\'esz\'aros, Rees, \& Wijers 1999), so an upper limit to the
total required formation rate would be 60\,GEM. We may have seen beaming
of about that factor or a bit less in GRB\,990123 (Kulkarni et~al. 1999),
but other bursts (e.g.\ 970228, 970508) show no evidence of beaming in
the afterglows (which may not exclude beaming of their gamma rays).
At present, therefore, any progenitor with a formation rate of
10\,GEM or more should be considered consistent with the observed GRB
rate.

   \section{Conclusions}
   \label{con}

We have shown that rapidly rotating black holes are an attractive power
source for gamma-ray bursts. Via the Blandford-Znajek mechanism (1977)
they can supply sufficient energy at a high rate. They also occur
often enough to explain the observed GRB rate, even if the gamma-ray
emission of a typical GRB is beamed to less 1\% of the sky. Because of
the requirement of rapid spin, the direct collapse of a stellar core
to a black hole is a less likely candidate for making GRB (at least
via the BZ effect). Therefore, mergers are much more attractive, which
implies a natural connection between GRBs and strong sources of gravity
waves. With advanced LIGO, the detection rate of mergers is predicted to
become large enough that direct verification of events that produce both
gravity wave and gamma-ray signals will become feasible, and will directly
constrain GRB beaming.
We have used the population synthesis calculations of Bethe \&
Brown (1998, 1999) to estimate the LIGO detection rate. We found it to
be dominated by black-hole, neutron-star mergers, and to be higher by a
factor 26 than previous estimates. As a result, we conclude that the most
energetic phenomena in astrophysics stem from black holes, whose defining
characteristic is paradoxically that no radiation can escape from them.

\acknowledgements

We would like to thank Roger Blandford, Chris Fryer, Sterl Phinney,
Simon Portegies Zwart, Kip Thorne and Stan Woosley for useful
suggestions and advice. 
This work was partially supported by the U.S. Department of Energy under 
Grant No.  DE--FG02--88ER40388.
HKL is also supported partly by  KOSEF 985-0200-001-2.

\newpage

\begin{table}
   \caption{Summary of the formation rates of
  various sources of gamma-ray bursts (GRB) or gravity waves (GW).
  L(H)BH means low- (high-)mass black hole.}
\begin{tabular}{@{}lccr@{}} \hline
object                               & GRB & GW & rate \\
                                     &     &    & [GEM$^1$] \\ \hline
NS - NS merger              &  X  & X  &  10 \\
NS - BH merger              &  X  & X  & 100 \\
WR star - LBH merger        &  X  &    & 380 \\ 
hypernova (HBH formation)   &  X  &    & 250 \\ 
BH in soft X-ray transient  &  X  &    &  90 \\ \hline 
\multicolumn{4}{@{}p{0.47\textwidth}@{}}{$^1$GEM means Galactic Events
        per Megayear; rates are quoted for redshift 0.}
\end{tabular}
\end{table}

\end{document}